\documentclass[a4paper]{jpconf}

\usepackage{graphicx}
\usepackage{amsmath,amssymb}
\usepackage{epstopdf}

\newcommand{\Eqref}[1]{Eq.~(\ref{#1})}
\newcommand{\nn}{\nonumber}
\newcommand{\be}{\begin{equation}}
\newcommand{\ee}{\end{equation}}
\newcommand{\bear}{\begin{eqnarray}}
\newcommand{\eear}{\end{eqnarray}}

\def \openone{\hbox{$1\hskip -1.2pt\vrule depth 0pt height 1.6ex width 0.7pt\vrule depth 0pt height 0.3pt width 0.12em$}}
\def \onlinecite #1{\cite{#1}}

\newcommand{\md}{\mathrm{d}}

\newcommand{\TuttiFrutti}{London:35,Gorter:49,Onsager,GL,London:50,LifshitzAzbelKaganov,
LL9,LL6,LL10,Abrikosov,Cabrera:89,Mishonov:94a,Basov:05,Plakida:10}
\newcommand{\Miro}{Miro:JPCS,Miro:19a,Miro:19b,Miro:BPU,Miro:20}

\begin{document}

\title{Scientific instrument for creation of effective\\ Cooper pair mass spectroscopy}

\author{Todor M. Mishonov and Albert M. Varonov}

\address{Georgi Nadjakov Institute of Solid State Physics, Bulgarian Academy of Sciences,
72 Tzarigradsko Chaussee Blvd., BG-1784 Sofia, Bulgaria}

\ead{mishonov@bgphysics.eu, varonov@issp.bas.bg}
\date{24 September 2020}

\begin{abstract}
We describe electronic instruments for creation of effective Cooper pair spectroscopy.
The suggested spectroscopy requires study of
electric field effects on the surface of cleaved superconductors.
The electronic instrument reacquires low noise amplifier with 
10$^6$ amplitude amplification 
which we have formerly used for study of Johnson-Nyquist and Schottky noises.
The nonspecific amplifier is followed by high-Q tunable resonance filter
based on schematics of general impedance converter topology which is also
and innovative device. 
The work of the device is based on the Manhattan equation of operational amplifier. 
After a final nonspecific amplification the total amplification can exceed 10$^9$
and in such a way sub-nano-volt signals can be reliably detected.
In short the observation of new effects in condensed matter physics 
leads to creation of new generation of electronic equipment.
\end{abstract}

\section{Introduction}
Effective masses $m^*$ of charge carriers are important notions of the physics of condensed 
matter.
In the physics of metals and semiconductors they represents basic electronic phenomena
related to thermodynamics and transport properties.
Alas the physics of superconductivity is an exception.
Except for some episodic hints, effective masses of super-fluid charge carriers have never been systematically determined.
The purpose of the present study is to evaluate experimental efforts necessary for the creation
of Cooper pair mass spectroscopy: equipment, samples, consumptives, and perspectives
for further development.
A review of present status of the problem together with detailed theory is given in 
our recent papers~\cite{m*_PhysC,m*_Scripta}
and references therein~(\cite{\TuttiFrutti} and so on). 
In the next Sec.~\ref{Theory} we recall the main theoretical results,
then we  describe the suggested set-ups in Sec.~\ref{Set-ups}.
In the Sec.~\ref{Electronics} we describe the electronic equipment which has to be build.
Later on in Sec.~\ref{Experiment} we describe the necessary samples 
and the experimental data processing for the planned first successful experiments.

\section{Theory Review}\label{Theory}
In this section we recall only the final theoretical formulae which have to be used 
for the creation  of Cooper pair mass spectroscopy. 

\subsection{Current induced contact potential difference}

The initial idea for creation of Cooper pair mass spectroscopy is to use current induced
contact potential difference or Bernoulli effect in superconductors. 
For the  detailed list of references, physics and history of the problem see our recent article~\onlinecite{m*_PhysC}.
We suppose capacitive coupling of the cleaved superconductor by 4 electrodes.
Between electrodes (1) and (3) is applied 
driving voltage, which is a sum of two sinusoidal:
one basic with frequency high $\Omega$ and 
another one modulated with much smaller frequency $\omega\ll\Omega$
\be
U_{1,3}(t)
=U_a\sin(\Omega t)+U_b\sin(\Omega t)\sin(\omega t)
=U_a\sin(\Omega t)
+\frac12 U_b\,[\cos((\Omega-\omega)t)-\cos((\Omega+\omega)t)].
\ee
Then between those probes (1) and (3) a current flows
\begin{align}&
I_{1,3}(t)=C_{1,3}\mathrm{d}_t U_{1,3}
=I_a\cos(\Omega t)-\frac12\,I_b\sin((\Omega-\omega)t)
+\frac12\,I_b\sin((\Omega+\omega)t),\\
&
I_a\approx\Omega\, C_{1,3} \,U_a,
\qquad 
I_b\approx\Omega\, C_{1,3}\, U_b,
\qquad C_{1,3}\equiv\frac{C_1C_3}{C_1+C_3},
\end{align}
where $C_1$ and $C_3$ are the capacitances between the electrodes and the superconductor samples.

This electrostatically driven current $I_{1,3}(t)$ in thin superconducting film with thickness
$d_\mathrm{film}\ll \lambda(T)$ much smaller than penetration depth $\lambda{T}$
creates the Bernoulli voltage
\begin{align}
&U_{2,4}(t)=-\mathcal BI_{1,3}^2
=-U_\mathrm{B}\sin(\omega t)+\dots,
\qquad
U_\mathrm{B}\equiv\frac12 \mathcal B I_aI_b.
\nonumber
\end{align}
where the high frequency terms with frequencies $2\Omega\pm\omega$
are not written.
The purpose of the experiment is to use the voltage amplitude $U_\mathrm{B}$ of the
demodulated signal with frequency $f=\omega/2\pi$.
The ratio 
\be
\mathcal B(T) \equiv U_\mathrm{B}/I_aI_b
\ee
of the experimentally measurable amplitudes 
determines the effective mass of Cooper pairs
\be
m^*
=\frac{e^*}{4\pi^2 \mathcal B}
\dfrac{\ln\dfrac{R_2}{r_2}}{R_2^2-r_2^2}
\left(\frac{\lambda(0)\lambda(T)}
{\varepsilon_0c^2\,d_\mathrm{film}}\right)^{2},
\label{CPD}
\ee
where we suppose that the second electrode is 
a ring internal and external radii $r_2$ and $R_2$.
The geometry is described in the next section.
\subsection{Electrostatically modulated thin film kinetic inductance}

Historically, the first experimental determination of effective mass of Cooper pairs
has been performed~\cite{Fiory:91,Reply} using the formula~\cite{Comment}
\be
m^*=-2|e|\mathrm{sgn}(e^*) L(0)L(T)\frac{\delta Q}{\delta L(T)},
\qquad
\mathcal{C}_s(T)=\frac{L(0)}{L(T)},
\label{inductance}
\ee
where $L(T)$ 
is the temperature dependent kinetic inductance per square sample,
$L(0)$ analogously to $\lambda(0)$ ts extrapolated to zero temperature value,
$\delta L(T)$ is the change of the kinetic inductance under
the electrostatic modulation by extra electric charge per unit area 
$\delta Q$, i.e. the electrostatic induction $D_z.$
\subsection{Electric field induced surface magnetization of the vortex phase}

Last possibility for creation of Cooper pair mass spectroscopy
is to study sinusoidal modulation of the excess surface magnetization
per unit area $M$ under the modulation of surface charge density $Q$
\begin{equation}
m^{*}=\dfrac{e^* B \lambda^2(0)}{4  (\mathrm{d} M/\mathrm{d} Q)} 
=\frac{e B \lambda^2(0) Q_0}{2\,\mathrm{sgn}(e^*) M_0}, \qquad
Q(t)=Q_0 \sin (\omega t),\qquad
M(t)=M_0 \sin (\omega t),
\label{magnetization}
\end{equation}
where $B$ is the external magnetic field which is perpendicular to the superconductor surface of extreme type-II superconductor which are, for example, all high-$T_c$ cuprates.

\section{Experimental Set-ups}\label{Set-ups}
\begin{figure}[h]
\centering
\includegraphics[scale=0.35]{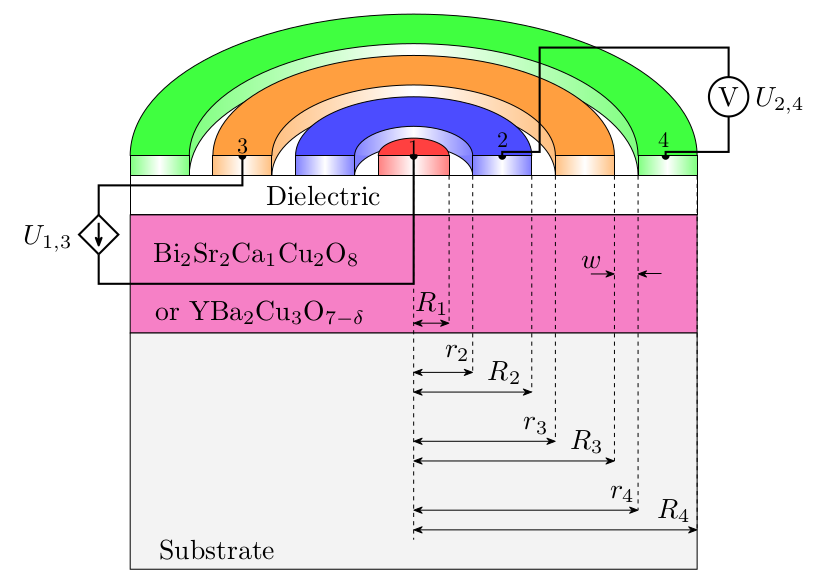}
\caption{
Experimental set-up for electrostatic excitation of current induced contact potential difference~\cite{m*_PhysC}.
Drive electrodes (1) and (3) and detector electrodes (2) and (4) are coaxial normal metal rings.
All four electrodes are plates of plane capacitors.
The other plane is the cleaved surface of a superconducting film.
The effective mass of cooper pairs is determined according to the Eq.~(\ref{CPD})
by voltages $U_{1,3}$ and $U_{2,4}.$
\label{Fig.:CPD}
}
\end{figure}

\begin{figure}[h]
\centering
\includegraphics[scale=0.72]{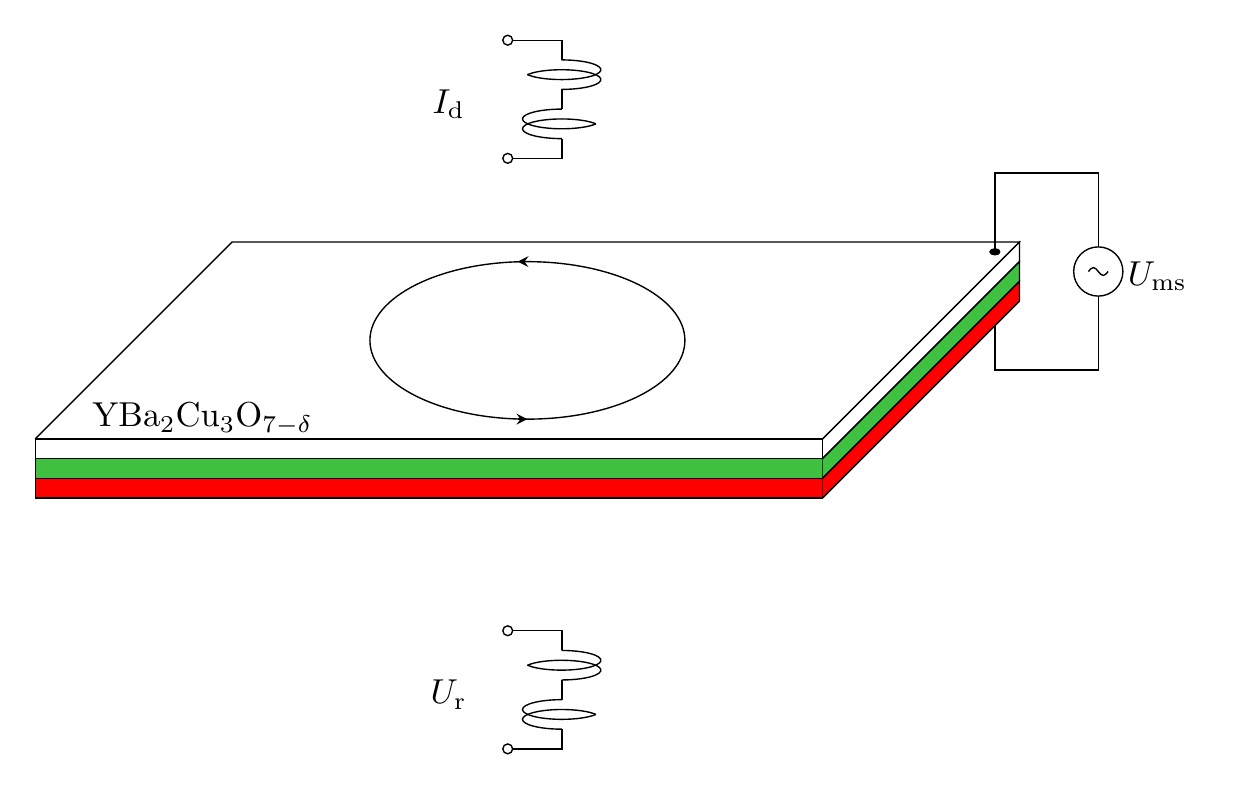}
\caption{Schematics (not to scale) representation of the experimental set-up for the
first measurement~\cite{Fiory:88,Fiory:91,Comment,Reply} of the effective mass of cooper pairs $m^*$
determined according to Eq.~(\ref{inductance}).
The eddy currents in the superconducting film are induced by the current through the drive
coil $I_\mathrm d$ and detected by the voltage $U_\mathrm r$ induced in the detector coil.
The modulating voltage $U_\mathrm{ms}$ changes the mutual inductance between coils
parameterized by the kinetic inductance per square sample $L(T)$.
}
\label{Fig:L_modulation}
\end{figure}

\begin{figure}[h]
\centering
\includegraphics[scale=0.35]{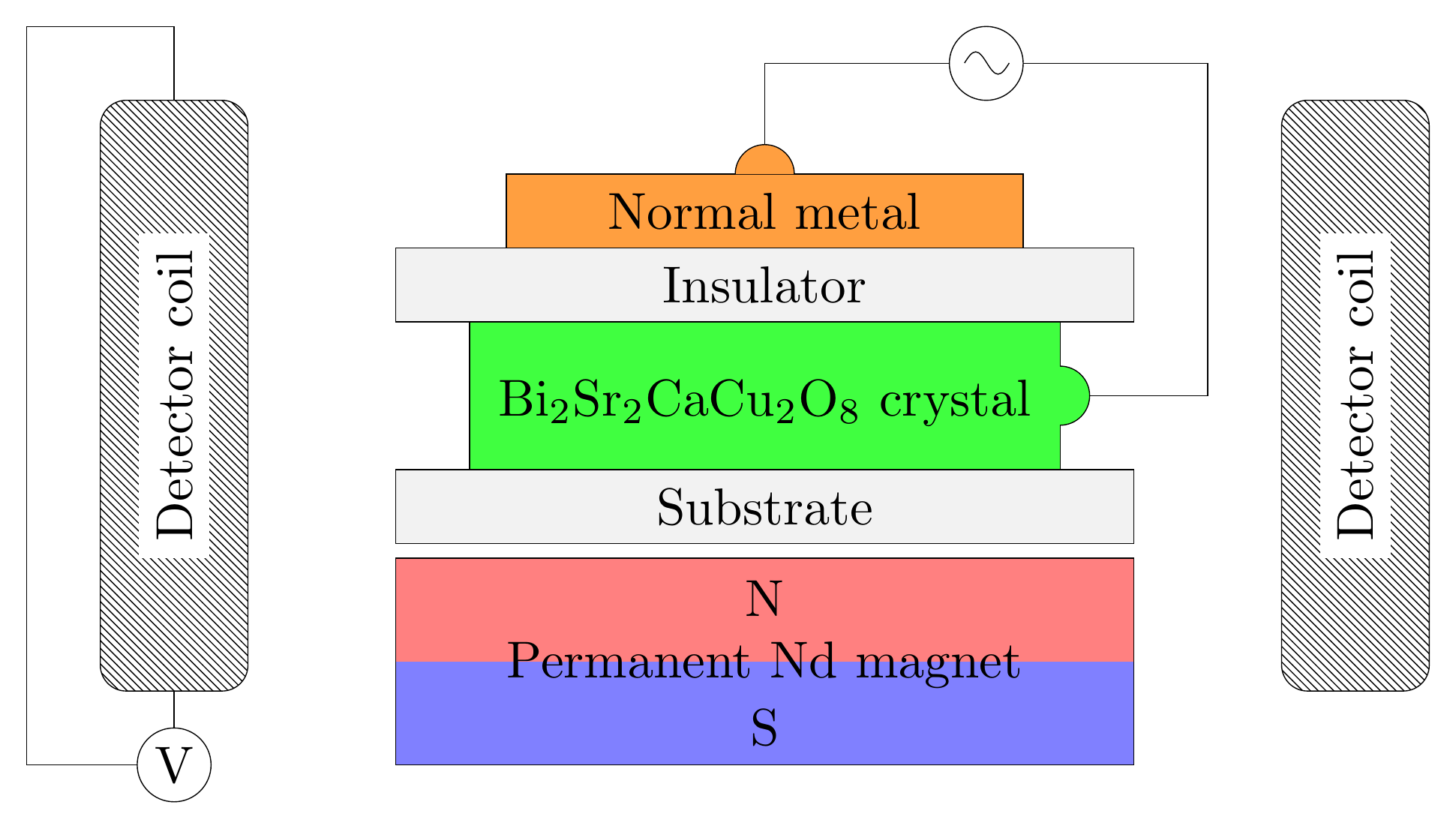}
\caption{Set-up for study of surface magnetization of the vortex phase of superconductors
induced by AC electric field applied to the surface~\cite{m*_Scripta}.
Magnetic field $B\ll B_{c2}(T)$ is created by a Nd permanent magnet.
Bi$_2$Sr$_2$Ca$_1$Cu$_2$O$_8$ crystal is grown on a substrate.
The cleaved upper surface of the crystal is one plate of a plane capacitor
with insulator layer. 
The other plate is a normal metal. 
An AC electric voltage is applied between the crystal and the normal electrode.
The theoretically predicted AC magnetization is measured 
as a voltage in the detector coil.
The new effect is parameterized by the effective mass of Cooper pairs $m^*$
according to Eq.~(\ref{magnetization}).
}
\label{Fig:Surf_Mag}
\end{figure}

Here we describe the set-ups for the experiments,
which theory we review in the former section.
The set-ups corresponding to the described 3 methods are depicted in 
Figs.~\ref{Fig.:CPD}, \ref{Fig:L_modulation} and \ref{Fig:Surf_Mag}

After this brief description of the theory and set-ups in the next section we will
focus on the electronic necessary to be build for the suggested Cooper pair mass spectroscopy.

\section{Electronics of the instrument for Cooper pair mass spectroscopy}\label{Electronics}

\begin{figure}[h]
\centering
\includegraphics[scale=0.45]{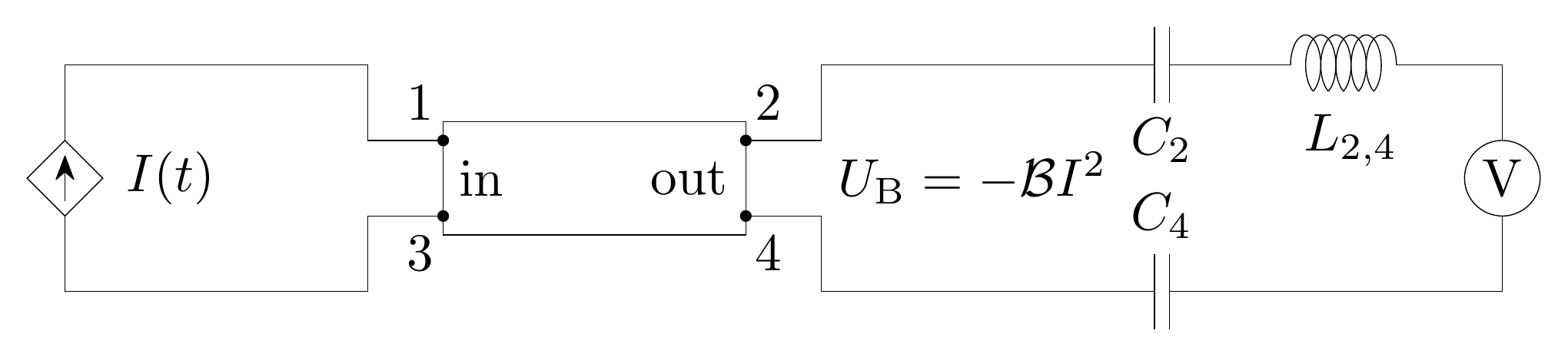}
\caption{Effective schematics for measurement of current induced contact potential difference. 
Capacitively induced current $I(t)$ creates CPD according Bernoulli theorem.
The AC CPD is measured after two capacitors with sequential capacitance $C_2 C_4/(C_2+C_4)$.
The pre-amplifier and the lock-in are represented symbolically as voltmeter.}
\label{Fig:Circ_}
\end{figure}

\begin{figure}[h]
\centering
\includegraphics[scale=0.3]{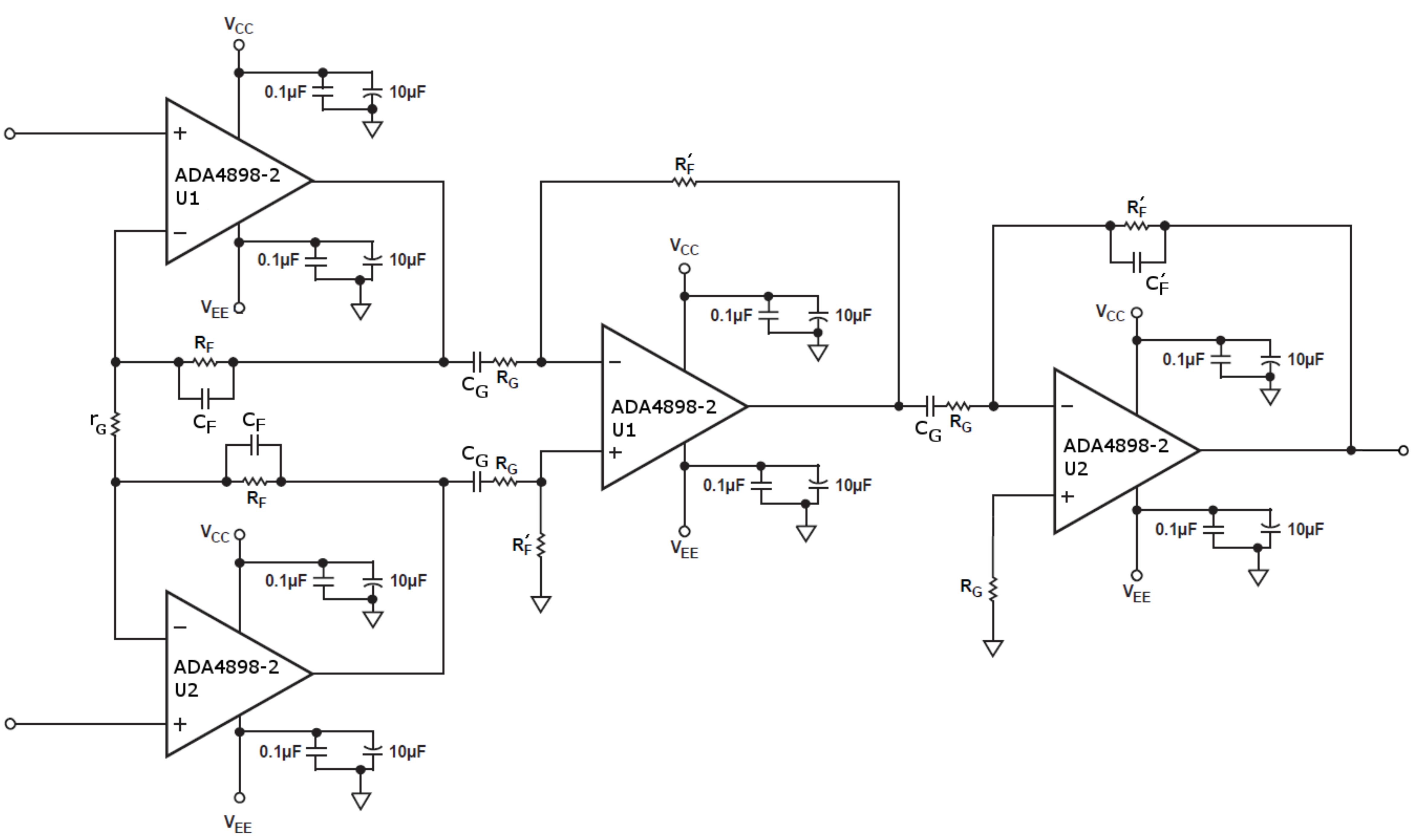}
\caption{A reliable pre-amplifier for measurements of CPD after Refs.~\cite{epo5,epo5:a}.
One can see a buffer and difference amplifier forming an instrumental amplifier
which is sequenced by an non-inverting amplifier.}
\label{Fig:Amplifier}
\end{figure}

All considered in the former section experiments require measurement of nano-volt range 
signals which is below the standard electronics equipment and requires development on unique
electronics.
For example the block-scheme for observation of current induced CPD is schematically
represented in Fig.~\ref{Fig:Circ_}.
The symbolically denoted there voltmeter starts with a pre-amplifier shown in Fig.~\ref{Fig:Amplifier}.
This pre-amplifier is reliable and we used it for education experiments
for determination of the Boltzmann constant~\cite{epo5}
and the electron charge~\cite{epo6}.
The device is so robust that we reproduced in more than 300 samples for the set-ups 
of 5-th and 6-th high-school Experimental physics olympiads~\cite{epo5:a,epo6:a}.
The pre-amplifier consists of a buffer with a double operational amplifiers. 
The buffer has a frequency dependence of the amplification as for a non-inverting amplifier
$Y_\mathrm{NIA}$. 
After the buffer we have a difference amplifier with frequency dependence of the
amplification coinciding with the last inverting amplifier $Y_{\Delta}=Y_\mathrm{IA}$.
In such a way the total amplification is given by the product
\be
\Upsilon (\omega)=\Upsilon_\mathrm{NIA}(\omega) \Upsilon_\Delta(\omega) \Upsilon_\mathrm{IA}(\omega) .
\label{sequential_amplification}
\ee
The offset voltages of the operational amplifiers are stopped by the large capacitors  
$C_\mathrm{G}$ followed by large gain resistors 
$R_\mathrm{G}$. 
In such a way for relatively low frequencies for which nevertheless 
$\omega R_\mathrm{G}C_\mathrm{G}\gg1$
the amplification is approximately constant
\be
\Upsilon(\omega \rightarrow 0) \equiv Y=Y_\mathrm{NIA} Y_{\Delta} Y_\mathrm{IA} =
\left(1+2\frac{R_\mathrm{F}}{r_\mathrm{_G}} \right )
\left(\frac{R_\mathrm{F}^\prime}{R_\mathrm{G}} \right)^{2} \nn
\ee
which can be clearly seen at the frequency dependence of the frequency
calculated according to our formulae~\cite{epo5}
\begin{align}
& G^{-1}(\omega)=G_0^{-1}+\mathrm{j}\omega\tau\ , \qquad  \tau \equiv 1/2\pi f_0 \\
& \Upsilon_\mathrm{NIA}(\omega)  = 
\frac1{G^{-1}(\omega)+y^{-1}(\omega)}, \qquad
y(\omega) \equiv \frac{Z_\mathrm{F}(\omega)}{r_\mathrm{_G}}+1, \qquad
\frac1{Z_\mathrm{F}(\omega)}=\frac1{R_\mathrm{F}}+\mathrm{j}\omega C_\mathrm{F} \nn \\
& \Upsilon_\Delta (\omega) =
\frac{-1}{\Lambda(\omega)+G^{-1}(\omega)[1+\Lambda(\omega)]}, \quad 
\Lambda(\omega) \equiv \frac{Z_\mathrm{G}(\omega)}{R_\mathrm{F}^\prime}, \qquad
Z_\mathrm{G}(\omega)=R_\mathrm{G}+\frac1{\mathrm{j} \omega C_\mathrm{G}}, \nn \\
& \Upsilon_\mathrm{IA} (\omega) =
-\frac{1}{\Gamma(\omega)+G^{-1}(\omega)[1+\Gamma(\omega)]}, \quad
 \Gamma(\omega) \equiv \frac{Z_\mathrm{G}(\omega)}{Z_\mathrm{F}^\prime}, \qquad
 \frac1{Z_\mathrm{F}^\prime(\omega)}=\frac1{R_\mathrm{F}^\prime}+\mathrm{j}\omega C_\mathrm{F}. \nn
\end{align}
and shown in Fig.~\ref{sequential_amplification}.
\begin{figure}[h]
\centering
\includegraphics[scale=0.6]{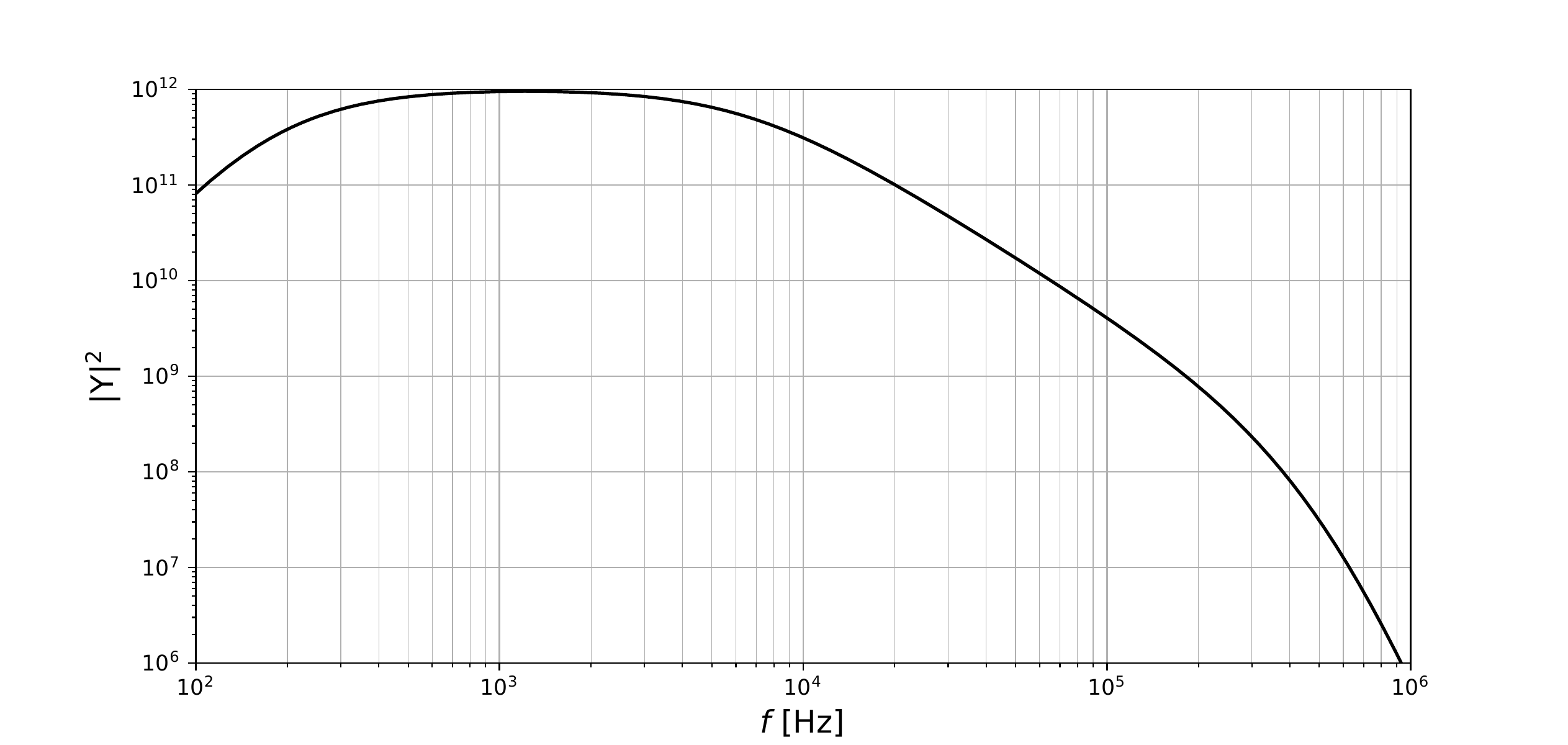}
\caption{Frequency dependence of the amplification of the schematics represented
in Fig.~\ref{Fig:Amplifier}. The theoretical formulae for the transmission of every block are 
given in Eq.~\ref{sequential_amplification}.}
\end{figure}
For this study we re-derived the Manhattan equation of OpAmps~\cite{Manhattan_AIP},
applied it in the theory of amplifiers~\cite{JPhysComm:19}, and studied 
statistical properties of the crossover frequency of the crossover frequency $f_0$
of the used low-noise OpAmp ADA4898-2~\cite{ADA4898}.

\begin{figure}[h]
\centering
\includegraphics[scale=0.4]{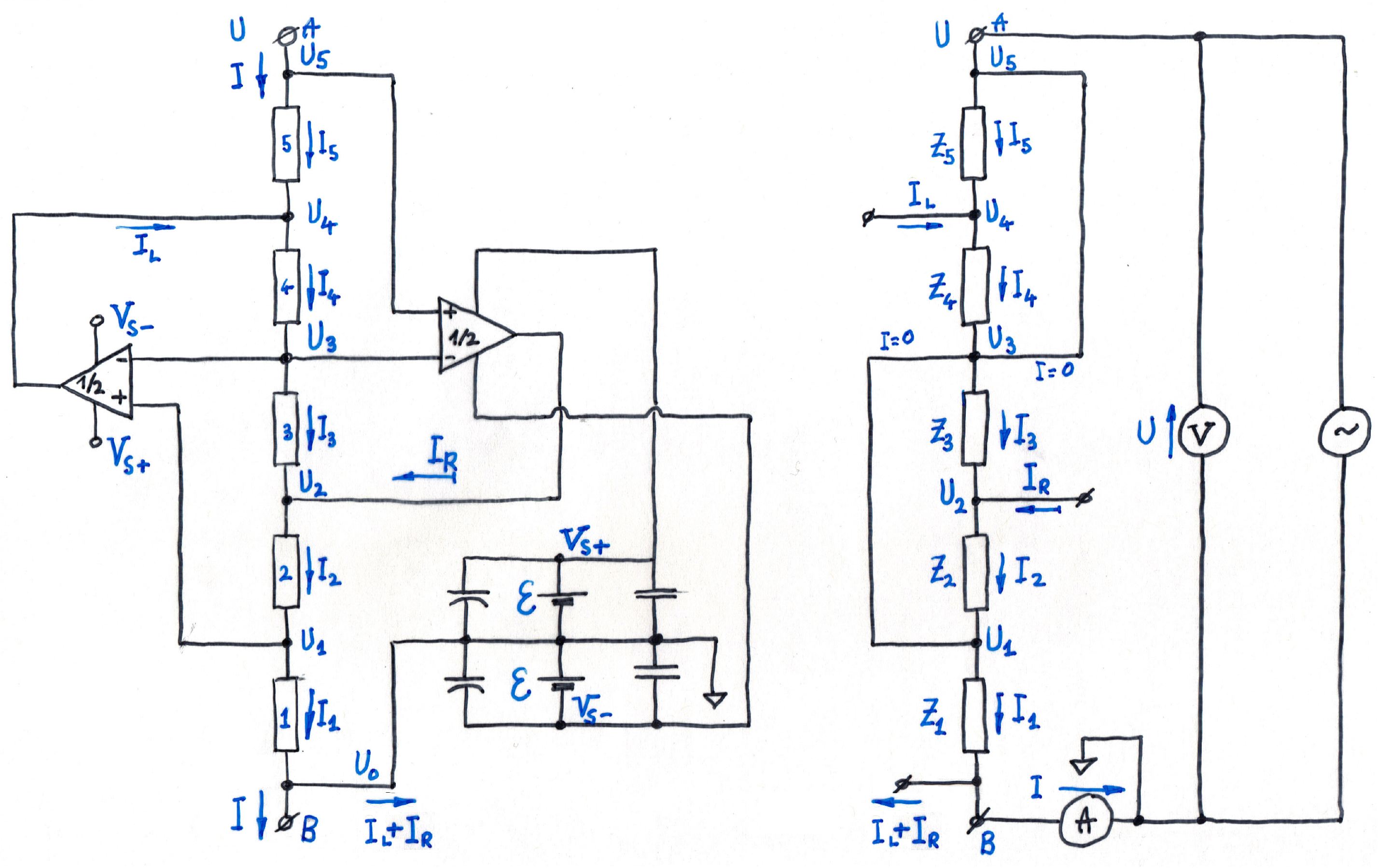}
\caption{Left: GIC patent CN86205529U~\cite{GIC:Pat,epo7:a}.
Right: Effective schematics for calculation of low frequency behavior 
when operation amplifier output currents equalize the input voltages.
General theoretical formula for the effective impedance of GIC is given by Eq.~\ref{GIC_impedance} and both experiment and theory are graphically represented in Fig.~\ref{Fig.:GIC_experiment}
\label{GIC_schematics}
}
\end{figure}

\begin{figure}[h]
\centering
\includegraphics[scale=0.7]{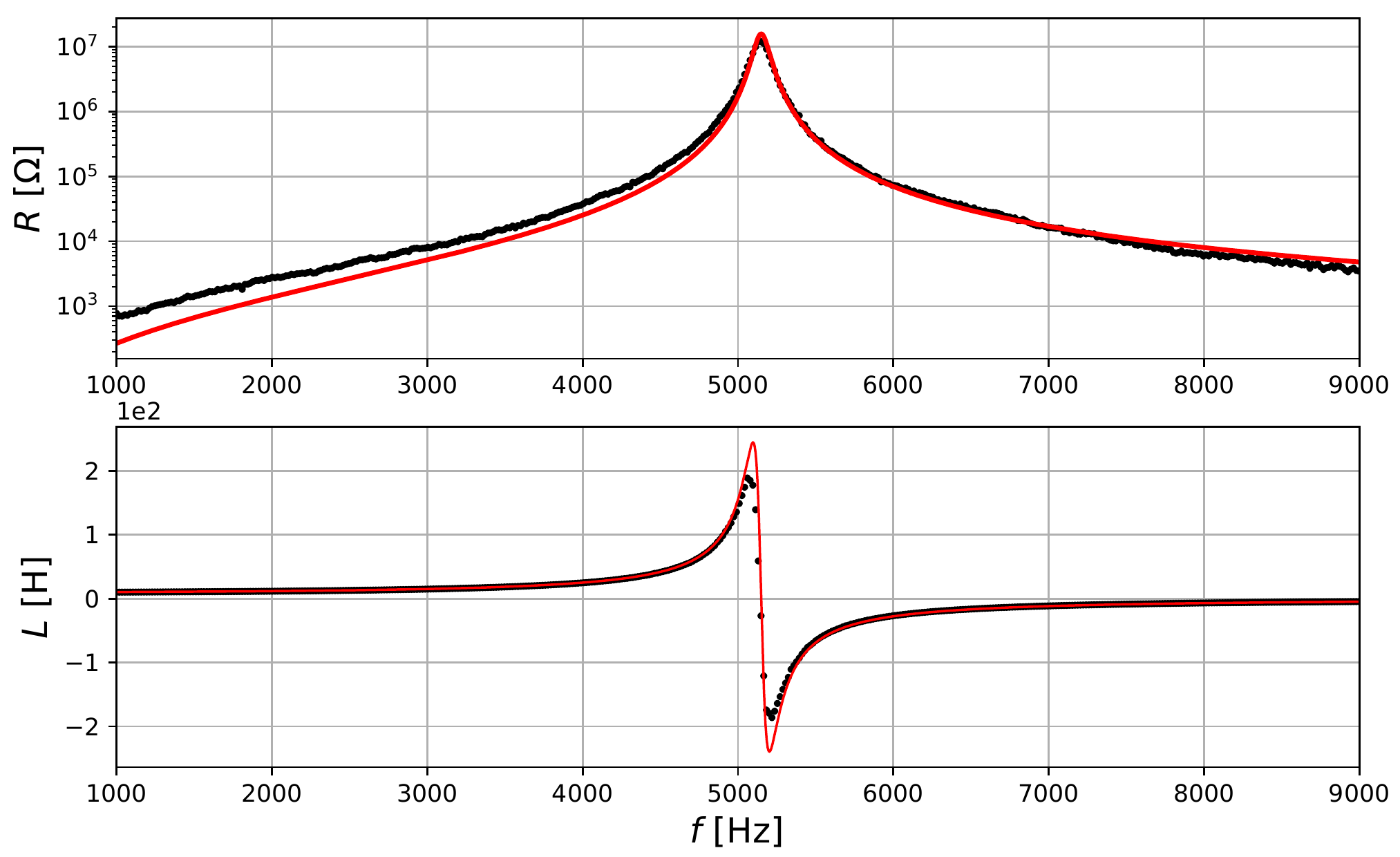}
\caption{Experimental data points and theoretical curves \Eqref{GIC_impedance}
for the frequency dependence of a resonator implemented by GIC.
The upper figure is the frequency dependence of the resistance $R(f)=\Re(Z)$,
while the lower one is the frequency dependence of the inductance $L(f)=\Im(Z)/\omega.$
At small frequencies we have an almost ideal inductance,
then we have a high-Q resonance and after it the inductance turns into capacitance.
The real part of the impedance, which is \textit{de facto} the resistance
is almost symmetric with respect to the resonance.
}
\label{Fig.:GIC_experiment}
\end{figure}

Concerning the nonspecific amplifier, we have reached saturation 
using one of the best low-noise OpAmp after 10$^6$ amplification
the signal is slightly below the overload regime.
In order to go further for observation of new physical effects we need to construct
a new type resonance filter having Q-factor comparable with quartz resonator
but having advantage to be tunable.
We have solved this long standing endeavor using 
the topology of general impedance converter (GIC)~\cite{GIC:Pat}
drawn in Fig.~\ref{GIC_schematics}.
The general theoretical formula we have derived ($\alpha, \, \beta \equiv G^{-1}(\omega)$)
\be
Z_{_\mathrm{GIC}}(\omega)=\dfrac{U}{I}
=\dfrac{Z_5}
{1-\dfrac{(Z_1+Z_2)-(Z_3+Z_4)K}{(1+\beta)(Z_1+Z_2)-Z_3K}}, \qquad
 K\equiv\frac{Z_2+ (Z_1+Z_2)\alpha}{Z_3+(Z_3+Z_4)\alpha},
\label{GIC_impedance}
\ee
perfectly match the experimental data depicted in Fig.~\ref{Fig.:GIC_experiment}.
The detailed description of the resonator will be published elsewhere~\cite{epo7:a}.
Now we can describe in short the planned experiment.

\section{Experiment}\label{Experiment}
All details of the planned experiment are already tested from the fundamental theory of 
a new effect in superconductors.
The last missing ling of the chain was the resonator by
GIC
and now we can start creation of the 
scientific instrument for the new Cooper pair mass spectroscopy.
In order to alleviate the success in the initial stage we recommend the use of 
90~K Bi$_2$Sr$_2$Ca$_1$Cu$_2$O$_8$ mono-crystal which can be easily cleaved.
In conclusion we believe that within one year we will have first intentionally done experiment 
for determination of $m^*$ in high-$T_c$ cuprates.

\section*{Acknowledgments} 
The authors thank for correspondence, suggestion and interest to the present project to 
Dmitri Basov, Hassan Chamati, Mauro Doria, 
Stefan Drechsler, Vadim Grinenko, Alexei Koshelev, 
Milorad Milosevic, Nikolay Plakida, Herman Suderow, 
Thomas Timusk, and Nikolay Tonchev.
The authors appreciate the friendly atmosphere of the school. 
After presenting our idea for measurements of surface magnetization 
during the discussion sections and afterwards,
we learned a lot about molecular magnets from Miroslav Georgiev,
where similarly to the physics of superconductivity, the magnetic field plays an essential role~\cite{\Miro}.
Continuous efforts of Hassan Chamati ensured high level ``bio-field'' and scientific level.
One of the authors, AMV acknowledges the support by 
National program ``Young scientists and postdoctoral researchers''
approved by DCM 577, 17.08.2018.

\section*{References}

\bibliography{MishonovISCMP2020}


\end{document}